\documentstyle[twocolumn,aps]{revtex}

\begin{document}
\draft
\title{Microscopic relaxation mechanisms and linear magnetization dynamics}
\author{Vladimir L. Safonov and H. Neal Bertram}
\address{Center for Magnetic Recording Research, \\
University of California - San Diego, \\
9500 Gilman Drive, La Jolla, CA 92093-0401}
\date{\today}
\maketitle

\begin{abstract}
Linear magnetization dynamics in the presense of a thermal bath is analyzed
for two general classes of microscopic damping mechanisms. The resulting
stochastic differential equations are always in the form of a damped
harmonic oscillator driven by a thermal field. The damping term contains
both the interaction mechanisms and the symmetry of the magnetic system.
Back transformation from the oscillator coordinates to the magnetization
variables results in a macroscopic tensor form of damping that reflects the
system anisotropy. Scalar Landau-Lifshitz-Gilbert damping term is valid only
for systems with axial symmetry. Analysis of FMR linewith measurements
versus frequency, temperature, and film thickness in NiFe films shows good
agreement with a combination of slow-relaxing impurity and magnon-electron
confluence processes.
\end{abstract}


\section{Introduction}

The study of classical magnetization dynamics of fine ferromagnetic
particles and thin films is of great interest in connection with the rapid
progress of nanotechnologies and high-density magnetic recording. The most
important research problems in this field include: (a) the problem of
magnetization reversal from the former equilibrium direction to a new one
under an applied reversal magnetic field, (b) magnetization escape over a
barrier from one equilibrium direction to another one under thermal
agitation, and (c) the effect of magnetization thermal fluctuations on the
noise characteristics of magnetoresistive devices. All these problems depend
on relaxation properties of the system, i.e., detailed mechanisms of
magnetization interaction with other degrees of freedom (spin waves, elastic
vibrations, conduction electrons, impurities, etc.) of material. Previously
substantial effort has been focused on the study of mostly phenomenological
magnetization relaxation (e.g., \cite{landau},\cite{gilbert},\cite{callen}, 
\cite{Patton},\cite{baryakhtar},\cite{Silvabook}).

The aim of this paper is to analyze linear magnetization dynamics in the
presence of a thermal bath for two general classes of microscopic damping
mechanisms. This analysis will be restricted to low-level excitations where
the magnetization remains close to equilibrium. Recently, magnetization
relaxation by direct coupling processes have been analyzed for the case of a
general anisotropic magnetic system \cite{BertSafJin}. The resulting
stochastic differential equation obtained was in the form of a simple damped
harmonic oscillator driven by a random thermal fluctuating field. The
coordinate system of this oscillator (or, normal mode) reflected the
magnetic anisotropy of the system. Although specific damping mechanisms were
not analyzed, the damping parameter was equivalent to the relaxation time.
Transformation back to magnetization variables yielded macroscopic dynamics
where the damping term reflected the system symmetry in a tensor form \cite
{SafTensor},\cite{BertJinSaf},\cite{SafBertNoise}.

Here this analysis is expanded to examine generally both direct and indirect
mechanisms for coupling to a thermal bath. In all cases a stochastic
differential equation in the form of a damped harmonic oscillator with
random forces is obtained (and the corresponding tensor form of damping, as
well). In addition, two specific relaxation mechanisms are examined in
detail: the direct process of magnon-electron confluence and the indirect
process of `slow-relaxing' impurity. It is shown that these two processes in
combination can explain experimental measurements of ferromagnetic resonance
(FMR) linewidth in permalloy thin films. Using reasonable estimates for the
parameters associated with these mechanisms, very good fitting is obtained
to the linewidth dependencies on temperature, frequency and film thickness.

A phenomenological tool to study magnetization relaxation, which has had
widespread use, is the Landau-Lifshitz equation \cite{landau} or, its
modification with Gilbert damping \cite{gilbert}. These (LLG) equations
conserve the absolute value of the magnetization ($\left| {\bf M}\right| =%
{\rm const}${\rm )} in a single domain. They are relatively simple and
therefore have been utilized for various calculations and micromagnetic
simulations. In particular, these equations have been used to analyze the
frequency dependence of FMR linewidth measurements in permalloy films (e.g., 
\cite{PattonWilts},\cite{PattonFraitWilts}). Although, this data can be fit
using the LLG equations (with a constant, phenomenological damping $\alpha $%
), this fact can not be considered as an experimental verification for
validity of the general nature of the LLG approach. As we show in detail,
the general tensor form of the macroscopic equation reduces to the LLG form
only for axially symmetric systems.

The paper is organized as follows. In the Sec. II we formulate the basic
gyromagnetic motion in a rotating frame and demonstrate that gyromagnetic
magnetization dynamics is equivalent to the motion of a harmonic oscillator.
Physics of a damped oscillator by direct coupling to a thermal bath is
reviewed and expanded in Sec. III. In Sec. IV the general mechanism of
indirect coupling is analyzed. An overview of magnetization dynamics with a
brief comparison to the phenomenological LLG form is given in Sec. V.
Experimental FMR data for NiFe films are analyzed in Sec. VI. It is shown
that relaxation processes due to magnetization interactions with
slow-relaxing impurities and conduction electrons fit\ the data well. A
discussion in given at the end of the paper.

\section{Formulation of basic gyromagnetic motion in rotating frame}

\subsection{Small magnetization motion}

Let us consider small-amplitude magnetization motions of a single-domain
ferromagnetic particle in the vicinity of equilibrium state ${\bf M=}M_{s}%
\widehat{{\bf z}}_{0}$. Here $\widehat{{\bf z}}_{0}$ is the unit vector in
the equilibrium direction and $M_{s}$ is the saturation magnetization. The
magnetization rotation around effective field in this case, in general, is
elliptical and the magnetic energy can be represented as a quadratic form
(e.g., \cite{SafTensor},\cite{BertJinSaf}):

\begin{equation}
{\cal E}/V=\frac{H_{x_{0}}}{2M_{s}}M_{x_{0}}^{2}+\frac{H_{y_{0}}}{2M_{s}}%
M_{y_{0}}^{2}.  \label{energy}
\end{equation}
Here $\widehat{{\bf x}}_{0}$ and $\widehat{{\bf y}}_{0}$ are the unit
orthogonal vectors in the plane perpendicular to the equilibrium direction. $%
H_{x_{0}}$ and $H_{y_{0}}$\ are positive fields, which include both
microscopic and shape anisotropies and the external magnetic field and $V$
is the particle volume (see Appendix I for an example).

\subsection{Normal mode approach}

We begin with a common normal mode approach (e.g., \cite{sparks},\cite{ABP}%
), where the small oscillations of the transverse magnetization coordinates
are first transformed to ``rotating'' complex variables by a linearized
Holstein-Primakoff transformation \cite{hopri}: 
\begin{eqnarray}
M_{x_{0}} &\simeq &-\sqrt{\hbar \gamma M_{s}/2V}(a^{\ast }+a),  \label{hp-a}
\\
M_{y_{0}} &\simeq &i\sqrt{\hbar \gamma M_{s}/2V}(a^{\ast }-a).  \nonumber
\end{eqnarray}
The magnetic energy ({\ref{energy}}) now can be rewritten in the quadratic
form:

\begin{equation}
{\cal E}/\hbar ={\cal A}a^{\ast }a+({\cal B}/2)(aa+a^{\ast }a^{\ast }),
\label{quadform}
\end{equation}
where ${\cal A}=\gamma (H_{x_{0}}+H_{y_{0}})/2$ and ${\cal B}=\gamma
(H_{x_{0}}-H_{y_{0}})/2$. This classical transformation yields a form,
common to quantum mechanics, where $\ a$ and $a^{\ast }$ correspond to
creation and annihilation operators so that in (\ref{quadform}) the term $%
a^{\ast }a=1-M_{z_{0}}/M_{s}$. Note that for an isotropic system where $%
H_{x_{0}}=H_{y_{0}}$, the term $B$\ vanishes.

The dynamic precession equations are given by

\begin{equation}
da/dt=-i{\cal A}a-i{\cal B}a^{\ast },\quad da^{\ast }/dt=i{\cal A}a^{\ast }+i%
{\cal B}a.  \label{eq-a}
\end{equation}
The mixed terms in (\ref{eq-a}) can be eliminated by the linear canonical
transformation: 
\begin{eqnarray}
a=uc+vc^{\ast }, &\quad &a^{\ast }=uc^{\ast }+vc,  \label{canonic} \\
u=\sqrt{\frac{{\cal A}+\omega _{0}}{2\omega _{0}}}, &\quad &v=-\frac{{\cal B}%
}{|{\cal B}|}\sqrt{\frac{{\cal A}-\omega _{0}}{2\omega _{0}}}.  \nonumber
\end{eqnarray}
The magnetization components in terms of $c$ and $c^{\ast }$ can be
expressed as

\begin{eqnarray}
M_{x_{0}} &\simeq &-\left( \hbar \gamma M_{s}/2V\right) ^{1/2}\left(
H_{y_{0}}/H_{x_{0}}\right) ^{1/4}(c^{\ast }+c),  \label{MinC} \\
M_{y_{0}} &\simeq &i\left( \hbar \gamma M_{s}/2V\right) ^{1/2}\left(
H_{x_{0}}/H_{y_{0}}\right) ^{1/4}(c^{\ast }-c).  \nonumber
\end{eqnarray}
With this transformation the energy in terms of the normal mode coordinates
is simply: 
\begin{equation}
{\cal E}/\hbar =\omega _{0}c^{\ast }c,  \label{ener}
\end{equation}
where

\begin{equation}
\omega _{0}=\sqrt{{\cal A}^{2}-{\cal B}^{2}}=\gamma \sqrt{H_{x_{0}}H_{y_{0}}}
\label{FMRfreq}
\end{equation}
is the ferromagnetic resonance (FMR) frequency. The dynamic equations in
terms of $c$ and $c^{\ast }$ are now independent:

\begin{equation}
dc/dt=-i\omega _{0}c,\quad dc^{\ast }/dt=i\omega _{0}c^{\ast }.
\label{c_equations}
\end{equation}
In order to derive a damped motion for this oscillator it is necessary to
consider the interaction with a thermal bath.

\section{Direct coupling to a thermal bath}

Linear relaxation can be most simply described in the thermal bath
approximation which represents coupling to a reservoir in thermodynamic
equilibrium. The thermal bath model has the following restriction: the
interaction with the thermal bath is assumed to be weak. Thus, dynamics of a
single-domain ferromagnetic particle in the zeroth (no thermal bath)
approximation represents the gyromagnetic rotation of magnetization around
an effective field. This means that the symmetry of the gyromagnetic motion
is dominant in the system and imposes some restrictions for damping and
fluctuations. Mathematical methods for a dynamic system interacting with a
thermal bath have been developed in detail in quantum optics (see, e.g., 
\cite{Lax0},\cite{lax},\cite{Louisell},\cite{scully}). Here we apply these
methods for magnetic particle dynamics.

\subsection{Direct linear coupling with a thermal bath}

We review \cite{BertSafJin} and consider classically a general linear
interaction of a harmonic oscillator (\ref{ener}) with a set of harmonic
oscillators that represent the thermal bath. The Hamiltonian for small
oscillations \{$a^{\ast }$, $a$\} (see, (\ref{quadform})) interacting with
harmonic oscillators \{$b_{k}^{\ast }$, $b_{k}$\} has the form:

\begin{eqnarray}
{\cal H}/\hbar &=&{\cal E}/\hbar +\sum_{k}\omega _{k}b_{k}^{\ast }b_{k}
\label{Hamiltonian} \\
&+&\sum_{k}[G_{k}(ab_{k}^{\ast }+\text{c.c.})+F_{k}(ab_{k}+\text{c.c.})]. 
\nonumber
\end{eqnarray}
The first term is the energy of the magnetic system (\ref{quadform}), the
second term gives the oscillations of the thermal bath with coordinates
(modes) $b_{k}$ and $b_{k}^{\ast }$, and the last term gives a direct
interaction coupling. The last term, if transformed back to the
magnetization variables, can be represented as a scalar product ${\bf M\cdot
H}_{th}$, where ${\bf H}_{th}$ is an effective field from the thermal bath.

Applying the transformation (\ref{canonic}), we can write the dynamic
equations for $c$\ and $b_{k}$\ as

\begin{eqnarray}
dc/dt &=&-i\omega _{0}c-i\sum_{k}(\widetilde{G}_{k}b_{k}+\widetilde{F}%
_{k}b_{k}^{\ast }),  \label{Eqc} \\
db_{k}/dt &=&-i\omega _{k}b_{k}-i\widetilde{G}_{k}c-i\widetilde{F}%
_{k}c^{\ast }.  \label{Eqb}
\end{eqnarray}
Here $\widetilde{G}_{k}=G_{k}u+F_{k}v$ and $\widetilde{F}_{k}=G_{k}v+F_{k}u$%
. The terms $\widetilde{F}_{k}b_{k}^{\ast }$ in (\ref{Eqc}) and $i\widetilde{%
F}_{k}c^{\ast }$ in (\ref{Eqb}) vanish as fast oscillating terms (they
describe rotations in the complex plane in the opposite direction to $c$ and 
$b_{k}$, respectively) and therefore can be omitted (see, \cite{BertSafJin}
for details). A solution of (\ref{Eqb}) is:

\begin{equation}
b_{k}(t)=b_{k}(0)e^{-i\omega _{k}t}-i\int_{0}^{t}\widetilde{G}%
_{k}c(t^{\prime })e^{-i\omega _{k}(t-t^{\prime })}dt^{\prime }.
\label{Eqbsol}
\end{equation}
Substituting (\ref{Eqbsol}) into (\ref{Eqc}), yields the
integro-differential equation for $c(t)$:

\begin{equation}
\frac{dc}{dt}=-i\omega _{0}c-\sum_{k}|\widetilde{G}_{k}|^{2}%
\int_{0}^{t}c(t^{\prime })e^{-i\omega _{k}(t-t^{\prime })}dt^{\prime }+f(t).
\label{slow}
\end{equation}
Here

\begin{equation}
f(t)=-i\sum_{k}\widetilde{G}_{k}b_{k}(0)e^{-i\omega _{k}t}
\label{whitenoise}
\end{equation}
describes a noise: $b_{k}(0)$ are random (and in thermal equilibrium). A
solution of (\ref{slow}) can be represented as $c(t)=\widetilde{c}(t)\exp
(-i\omega _{0}t)$, where $\widetilde{c}(t)$ is a ``slow'' variable.
Neglecting the memory for the slow variable ($\widetilde{c}(t^{\prime
})\rightarrow \widetilde{c}(t)$), we can rewrite Eq.(\ref{slow}) as:

\begin{equation}
\frac{dc}{dt}=-i\omega _{0}c-c\sum_{k}|\widetilde{G}_{k}|^{2}%
\int_{0}^{t}e^{i(\omega _{0}-\omega _{k})(t-t^{\prime })}dt^{\prime }+f(t).
\label{slow1}
\end{equation}
For a long time $t\gg 1/(\omega _{0}-\omega _{k})$ one can put the upper
integral limit at $\infty $ and use the formula:

\begin{equation}
\int_{0}^{\infty }du\exp (i\Omega u)=\pi \delta (\Omega )+i{\rm P.v.}\left( 
\frac{1}{\Omega }\right) ,  \label{DiracPV}
\end{equation}
where $\delta (\Omega )$ is the Dirac delta function and ${\rm P.v.}$ is the
principal value of the integral. Thus, we obtain the dynamic equation of a
damped harmonic oscillator:

\begin{equation}
dc/dt+\eta c=-i(\omega _{0}+\Delta \omega )c+f(t),  \label{principalOSC}
\end{equation}
where

\begin{eqnarray}
\eta &=&\pi \sum_{k}|\widetilde{G}_{k}|^{2}\delta (\omega _{0}-\omega _{k}),
\label{relratelinear} \\
\Delta \omega &=&-{\rm P.v.}\sum_{k}\frac{|\widetilde{G}_{k}|^{2}}{\omega
_{0}-\omega _{k}}.  \nonumber
\end{eqnarray}
The damping parameter $\eta $ (\ref{relratelinear}) represent a particular
form of ``Fermi's Golden Rule'' (e.g., \cite{sparks}). This form, for
example, has been used for calculating two-magnon scattering \cite{twomagnon}%
,\cite{AriasMills}\ and magnon-phonon \cite{mcmichael}\ processes.

\subsection{Arbitrary coupling to a thermal bath}

Lax \cite{Lax0},\cite{lax} has developed the most general approach for a
dynamic description of a harmonic oscillator interacting directly with a
thermal bath (reservoir). He used the following Hamiltonian: 
\begin{equation}
{\cal H}_{total}=\hbar \omega _{0}\widehat{c}^{\dag }\widehat{c}+(\widehat{c}%
\widehat{F}^{\dag }+\widehat{c}^{\dag }\widehat{F})+{\cal H}_{R},
\label{LaxHam}
\end{equation}
where $\widehat{c}^{\dag }$ and $\widehat{c}$ are creation and annihilation
Bose operators, $\widehat{F}$ and its Hermitian conjugate $\widehat{F}^{\dag
}$ describe thermal bath variables, ${\cal H}_{R}$ is the reservoir
Hamiltonian. Taking the density matrix equation (i.e., one of the most
fundamental equations in physics), Lax derived the master equation in the
second approximation on interaction and finally showed that the damped
harmonic oscillator should obey the following classical Langevin equation:

\begin{equation}
dc/dt+\eta c=-i(\omega _{0}+\Delta \omega )c+f(t).  \label{LaxHarmonic}
\end{equation}
Here $c=\langle \widehat{c}\rangle $ is the complex amplitude (classical
analog of annihilation operator) and $f(t)$ is a noise. The damping $\eta $
and frequency shift $\Delta \omega $ are defined by

\begin{equation}
\eta -i\Delta \omega \equiv -\frac{1}{{\hbar ^{2}}}\int\limits_{0}^{\infty
}du{\ }e{^{-i\omega _{0}u}}\langle \lbrack \widehat{F}(0),\widehat{F}^{\dag
}(u)]\rangle _{R},  \label{LaxRelax}
\end{equation}
where $\langle ...\rangle _{R}$\ denotes averaging over reservoir. As before
the formula (\ref{LaxRelax}) for elementary processes can be reduced to the
``Fermi's Golden Rule'' (second order perturbation theory, e.g., .\cite
{sparks}).

\subsubsection{Magnon-electron scattering}

As an example of magnetization relaxation due to direct coupling to the
thermal bath we shall consider the interaction of magnons with conduction
electrons in a ferromagnetic metal. From a microscopic point of view the
most probable process is the confluence of a magnon with wave vector ${\bf k}%
=0$ and energy $\hbar \omega _{0}$ and a conduction electron with wave
vector ${\bf k}_{1}\neq 0$ and energy $\epsilon ({\bf k}_{1})$ into a
conduction electron with wave vector ${\bf k}_{2}$ and energy $\epsilon (%
{\bf k}_{2})=\epsilon ({\bf k}_{1})+\hbar \omega _{0}$. It is obvious that
this process is forbidden for an ideal crystal, where the momentum
conservation is valid ${\bf k}_{2}={\bf k}_{1}+{\bf 0}$. Such confluence
processes exist only for non-uniform magnetic modes and demonstrate a strong 
$k$-dependence for the relaxation rate ($\propto k^{2}$) \cite{reizer},\cite
{solontsov}.

The abovementioned confluence process can occur in the presence of defects,
impurities or fluctuations which permit violation of momentum conservation
in the crystal (Fig.1a). This opportunity was qualitatively discussed by
Kambersky and Patton \cite{KamberskyPatton}. Because of its importance, we
shall consider this process in detail (see, Appendix B). The relaxation rate
for the process can be written as:

\begin{eqnarray}
\eta _{m-e} &=&c_{def}(u^{2}+v^{2})\omega _{0}\alpha _{c},  \label{kampat} \\
\alpha _{c} &\simeq &\frac{\pi }{N^{2}}\sum_{{\bf k}}\left( -\frac{\partial 
\overline{n}_{{\bf k}}}{\partial \omega _{e{\bf k}}}\right) \sum_{{\bf k}%
^{\prime }}|f_{{\bf kk}^{\prime }}|^{2}\delta (\omega _{e{\bf k}^{\prime
}}-\omega _{e{\bf k}}).  \nonumber
\end{eqnarray}
This relaxation rate is linear with the defect concentration $c_{def}$ and
frequency dependent: it contains transformation terms $(u^{2}+v^{2})\omega
_{0}=\gamma (H_{x_{0}}+H_{y_{0}})/2.$ The temperature dependence can occur
through the term: $\sum_{{\bf k}}\left( -\partial \overline{n}_{{\bf k}%
}/\partial \omega _{e{\bf k}}\right) $. An estimate of Eq.(\ref{kampat}) for 
$f_{{\bf kk}^{\prime }}=f={\rm const}$ and $T\ll \epsilon _{F}/k_{B}\sim
3\cdot 10^{4}$ K gives

\begin{equation}
\eta _{m-e}\simeq \frac{c_{def}}{16\pi ^{3}}\gamma (H_{x_{0}}+H_{y_{0}})%
\frac{m^{3}V_{0}^{2}|f|^{2}\epsilon _{F}}{\hbar ^{4}},
\label{estimatekampat}
\end{equation}
where $m$ is the conduction electron mass and $\epsilon _{F}$ is the Fermi
energy.

\section{Relaxation via an intermediate damped dynamic system}

There is a whole class of relaxation mechanisms that can not be analyzed as
elementary processes. The energy loss in this case occurs via intermediate
damped dynamic system. Here we shall consider the magnetization damping via
so-called, `slow relaxing' impurities (see, Fig.1b and Refs. \cite{mikhailov}%
,\cite{saffar},\cite{safbertimpurity},\cite{gurevich}). In this mechanism
the magnetization motion modulates the impurity splitting (levels, see,
Fig.1b). Thus the thermal equilibrium population of the energy levels varies
and transitions between the levels occur (arrows in Fig.1b). There is a
delay for these transitions due to a finite impurity relaxation time. This
delay results in a magnetization oscillation energy loss.

A general Hamiltonian describing the impurity level modulation is given in
Appendix C, Eq.(\ref{imp-ex}). In the case of only the coherent motion of
the host spins we have ${\bf S}({\bf R}_{j}+{\bf r}_{\nu })=-(V_{0}/\hbar
\gamma ){\bf M}$, where $V_{0}=V/N$ is the volume of elementary cell. Let us
consider a two-level paramagnetic impurity with the energy

\begin{equation}
{\cal H}_{imp,j}=\hbar \left[ \Omega _{0,j}+\delta \Omega _{j}(t)\right]
n_{j},  \label{hamimpu}
\end{equation}
where $\Omega _{0,j}$ is the splitting frequency, $n_{j}=s_{z_{j}}({\bf R}%
_{j})+1/2$ is the upper lever population and $j$ is the impurity index.
Applying Eqs.(\ref{MinC}) to (\ref{imp-ex}), we can write the impurity level
modulation as

\begin{equation}
\delta \Omega _{j}(t)=\Phi _{j}c(t)+\Phi _{j}^{\ast }c^{\ast }(t),
\label{implevelmod}
\end{equation}
where

\begin{eqnarray}
\Phi _{j} &=&\frac{V_{0}}{\hbar }\left( \frac{M_{s}}{\hbar \gamma V}\right)
^{1/2}\sum\limits_{\nu }\Big[B_{z_{j},x_{0}}({\bf R}_{j},{\bf r}_{\nu
})\left( \frac{H_{y_{0}}}{H_{x_{0}}}\right) ^{1/4}  \nonumber \\
&&+iB_{z_{j},y_{0}}({\bf R}_{j},{\bf r}_{\nu })\left( \frac{H_{x_{0}}}{%
H_{y_{0}}}\right) ^{1/4}\Big].  \label{amplitudeF}
\end{eqnarray}

The kinetics of the impurity population is defined by the following equation:

\begin{equation}
dn_{j}/dt=-\Gamma _{\parallel ,j}[n_{j}-n_{T}(\Omega _{j})].
\label{impurity}
\end{equation}
Here $\Gamma _{\parallel ,j}$ is the impurity relaxation rate and $%
n_{T}(\Omega _{j})=[\exp (\hbar \Omega _{j}/k_{B}T)+1]^{-1}$ is the
equilibrium population at frequency $\Omega _{j}=\Omega _{0,j}+\delta \Omega
_{j}(t)$. Taking into account 
\begin{eqnarray}
n_{j}(t) &=&n_{T}(\Omega _{0,j})+\delta n_{j}(t),  \label{dnont} \\
n_{T}(\Omega _{j}) &=&n_{T}(\Omega _{0,j})+\left[ \partial n_{T}/\partial
\Omega _{0,j}\right] \delta \Omega _{j}(t),  \nonumber
\end{eqnarray}
we can solve Eq.(\ref{impurity}) and obtain

\begin{equation}
\delta n_{j}(t)=\Gamma _{\parallel ,j}\frac{\partial n_{T}}{\partial \Omega
_{0,j}}\left[ \frac{\Phi _{j}c(t)}{\Gamma _{\parallel ,j}-i\omega _{0}}+%
\frac{\Phi _{j}^{\ast }c^{\ast }(t)}{\Gamma _{\parallel ,j}+i\omega _{0}}%
\right] .  \label{deltansol}
\end{equation}

The dynamic equation for the normal mode interacting with impurities is
obtained by substituting (\ref{implevelmod}) into (\ref{hamimpu}). This
yields an additional term to the basic mode Hamiltonian (\ref{ener}).
Similar to (\ref{c_equations}), the dynamic equation is now:

\begin{eqnarray}
dc/dt &=&-i\omega _{0}c-i\partial ({\cal H}_{imp,j}/\hbar )/dc^{\ast }
\label{dyneqnmi} \\
&=&-i\omega _{0}c-i\sum_{j}\Phi _{j}^{\ast }\delta n_{j}(t).  \nonumber
\end{eqnarray}
Substituting Eq.(\ref{deltansol}) into (\ref{dnont}) and (\ref{dyneqnmi}),
we obtain the equation of damped harmonic oscillator \cite{slownoise}:

\begin{equation}
dc/dt+\eta _{sr}c=-i(\omega _{0}+\Delta \omega _{sr})c+f(t).
\label{sr-oscillator}
\end{equation}
Here the frequency shift is equal to

\begin{equation}
\Delta \omega _{sr}\simeq -\sum\limits_{j}\left| \Phi _{j}\right| ^{2}\left(
-\frac{\partial n_{T}(\Omega _{0,j})}{\partial \Omega _{0,j}}\right) \frac{%
\Gamma _{\parallel ,j}^{2}}{\Gamma _{\parallel ,j}^{2}+\omega _{0}^{2}}
\label{freqshiftslow}
\end{equation}
and the relaxation rate is:

\begin{equation}
\eta _{sr}\simeq \sum\limits_{j}\left| \Phi _{j}\right| ^{2}\left( -\frac{%
\partial n_{T}(\Omega _{0,j})}{\partial \Omega _{0,j}}\right) \frac{\omega
_{0}\Gamma _{\parallel ,j}}{\Gamma _{\parallel ,j}^{2}+\omega _{0}^{2}},
\label{SlowRelaxRate}
\end{equation}
where

\begin{eqnarray}
\left| \Phi _{j}\right| ^{2} &=&\frac{M_{s}V_{0}}{\hbar \gamma N}\left[
\sum\limits_{\nu }B_{z_{j},x_{0}}({\bf R}_{j},{\bf r}_{\nu })\right]
^{2}\left( \frac{H_{y_{0}}}{H_{x_{0}}}\right) ^{1/2}  \nonumber \\
&&+\left[ \sum\limits_{\nu }B_{z_{j},y_{0}}({\bf R}_{j},{\bf r}_{\nu })%
\right] ^{2}\left( \frac{H_{x_{0}}}{H_{y_{0}}}\right) ^{1/2}.
\label{amplitImp}
\end{eqnarray}

To simplify (\ref{SlowRelaxRate}), we assume that $\Omega _{0,j}=\Omega _{0}$
and $\Gamma _{\parallel ,j}=\Gamma _{\parallel }$ for all impurities.
Summing over $j$ gives an average anisotropic exchange $B$:

\begin{eqnarray}
&&\sum\limits_{j}\left[ \sum\limits_{\nu }B_{z_{j},x_{0}}({\bf R}_{j},{\bf r}%
_{\nu })\right] ^{2}  \label{averaging} \\
&\simeq &\sum\limits_{j}\left[ \sum\limits_{\nu }B_{z_{j},y_{0}}({\bf R}_{j},%
{\bf r}_{\nu })\right] ^{2}=N_{imp}(z_{imp}B)^{2},  \nonumber
\end{eqnarray}
where $N_{imp}$ is the total number of impurities in the sample and $z_{imp}$
is the average number of magnetic neighbors for one impurity. Finally the
relaxation rate (\ref{SlowRelaxRate}) is:

\begin{eqnarray}
\eta _{sr} &\simeq &c_{imp}S\frac{(z_{imp}B)}{\hbar k_{B}T}^{2}\frac{\exp
(\hbar \Omega _{0}/k_{B}T)}{[\exp (\hbar \Omega _{0}/k_{B}T)+1]^{2}}
\label{final} \\
&&\times \frac{\gamma (H_{x_{0}}+H_{y_{0}})\Gamma _{\parallel }}{\Gamma
_{\parallel }^{2}+\omega _{0}^{2}},  \nonumber
\end{eqnarray}
where $c_{imp}=N_{imp}/N$ is the impurity concentration and $%
S=M_{s}V_{0}/\hbar \gamma $ is the value of the host spin. This relaxation
rate exhibits a definite temperature dependance. As discussed in the
previous section for magnon-electron scattering processes, the impurity
relaxation rate also is frequency dependent, via the term $\gamma
(H_{x_{0}}+H_{y_{0}})$ (with identical dependence as in magnon-electron
scattering).

\section{Macroscopic dynamic equations}

\subsection{Tensor form of magnetization damping}

We have seen that both direct and indirect magnetization interactions with a
thermal bath yield damped harmonic oscillator dynamics in the form: 
\begin{equation}
dc/dt+\eta c=-i(\omega _{0}+\Delta \omega )c+f(t).  \label{genharm}
\end{equation}
As discussed in \cite{SafTensor},\cite{BertJinSaf}, utilizing back
transformation (\ref{MinC}), from (\ref{genharm}) we can derive linearized
equations for the transverse magnetization components:

\begin{eqnarray}
\frac{d}{dt}\left( 
\begin{array}{c}
M_{x_{0}} \\ 
M_{y_{0}}
\end{array}
\right)  &=&\left( 
\begin{array}{cc}
-\eta  & -\gamma H_{y_{0}} \\ 
\gamma H_{x_{0}} & -\eta 
\end{array}
\right) \left( 
\begin{array}{c}
M_{x_{0}} \\ 
M_{y_{0}}
\end{array}
\right)   \label{mxmyFirst} \\
&&+\gamma M_{s}\left( 
\begin{array}{c}
h_{y_{0}}(t) \\ 
-h_{x_{0}}(t)
\end{array}
\right) .  \nonumber
\end{eqnarray}
Here

\begin{eqnarray}
h_{y_{0}}(t) &=&-\left( \frac{\hbar }{2\gamma M_{s}V}\right) ^{1/2}\left( 
\frac{H_{y_{0}}}{H_{x_{0}}}\right) ^{1/4}\left[ f^{\ast }(t)+f(t)\right] ,
\label{ranfields} \\
h_{x_{0}}(t) &=&-i\left( \frac{\hbar }{2\gamma M_{s}V}\right) ^{1/2}\left( 
\frac{H_{x_{0}}}{H_{y_{0}}}\right) ^{1/4}\left[ f^{\ast }(t)-f(t)\right] 
\nonumber
\end{eqnarray}
are the independent random fields.

The nonlinear macroscopic dynamic equation for an anisotropic magnetic
system which conserves the length of magnetization ($|{\bf M}|=M_{{\rm s}}$)
has the form \cite{SafTensor},\cite{BertJinSaf},\cite{SafBertNoise}:

\begin{equation}
\frac{d{\bf M}}{dt}=-\gamma {\bf M}\times \left[ {\bf H}_{{\rm eff}}+{\bf h}%
(t)\right] -\gamma \frac{{\bf M}}{M_{{\rm s}}}\times \lbrack \stackrel{%
\leftrightarrow }{\alpha }\cdot ({\bf M}\times {\bf H}_{{\rm eff}})].
\label{LL}
\end{equation}
Here $\stackrel{\leftrightarrow }{\alpha }$ is a dimensionless damping
tensor, which contains all necessary information about the system symmetry.
In the vicinity of equilibrium one has

\begin{equation}
\stackrel{\leftrightarrow }{\alpha }=\frac{\eta }{\gamma }\left( 
\begin{array}{ccc}
1/H_{y_{0}} & 0 & 0 \\ 
0 & 1/H_{x_{0}} & 0 \\ 
0 & 0 & 0
\end{array}
\right)  \label{damptensor}
\end{equation}
and Eq.(\ref{LL}) is equivalent to (\ref{mxmyFirst}).

Note that the tensor form of damping in (\ref{LL}) is a natural description
of an anisotropic system. It reflects an elliptical motion of magnetization
and contains just one relaxation parameter $\eta $. The $\stackrel{%
\leftrightarrow }{\alpha }$ should not be confused with a damping tensor
which contains two or more independent relaxation parameters.

\subsection{Linearized Landau-Lifshitz equation}

The Landau-Lifshitz equation with random fields ${\bf h}(t)$\ has the form:

\begin{equation}
\frac{d{\bf M}}{dt}=-\gamma {\bf M}\times \left[ {\bf H}_{{\rm eff}}+{\bf h}%
(t)\right] -\frac{\alpha \gamma }{M_{s}}{\bf M}\times \left( {\bf M}\times 
{\bf H}_{{\rm eff}}\right) ,  \label{principal}
\end{equation}
where $\alpha $ is a dimensionless damping parameter. Using (\ref{energy}),
we can calculate the effective field ${\bf H}_{{\rm eff}}=-\partial ({\cal E}%
/V)/\partial {\bf M}$ and write down the linearized equations for the
transverse magnetization components ($M_{z_{0}}\simeq M_{{\rm s}}$):

\begin{eqnarray}
\frac{d}{dt}\left( 
\begin{array}{c}
M_{x_{0}} \\ 
M_{y_{0}}
\end{array}
\right)  &=&\left( 
\begin{array}{cc}
-\alpha \gamma H_{x_{0}} & -\gamma H_{y_{0}} \\ 
\gamma H_{x_{0}} & -\alpha \gamma H_{y_{0}}
\end{array}
\right) \left( 
\begin{array}{c}
M_{x_{0}} \\ 
M_{y_{0}}
\end{array}
\right)   \label{mxmyLLG} \\
&&+\gamma M_{s}\left( 
\begin{array}{c}
h_{y_{0}}(t) \\ 
-h_{x_{0}}(t)
\end{array}
\right) .  \nonumber
\end{eqnarray}
We see that the non-diagonal terms in (\ref{mxmyLLG}) and (\ref{mxmyFirst}),
as expected, coincide with each other, respectively. The diagonal terms,
responsible for relaxation, in general, are different ($H_{x_{0}}\neq
H_{y_{0}}$). These damping terms are equal only in the special case when $%
H_{x_{0}}=H_{y_{0}}$ (see, Appendix A).

The Eq.(\ref{mxmyLLG}) can be rewritten in terms of normal modes using (\ref
{MinC}) as

\begin{equation}
dc/dt+\eta c+\widetilde{\eta }c^{\ast }=-i\omega _{0}c+f(t),  \label{LinLL}
\end{equation}
where

\begin{equation}
\eta =\alpha \gamma (H_{x_{0}}+H_{y_{0}})/2,\quad \widetilde{\eta }=\alpha
\gamma (H_{x_{0}}-H_{y_{0}})/2.  \label{LinLLdamp}
\end{equation}
We see that Eq. (\ref{LinLL}) contains one extra fast oscillating term $%
\widetilde{\eta }c^{\ast }$. As shown in Ref. \cite{BertJinSaf}, this term
changes the power spectral shape. However, in the case when the thermal term
is negligible, the fast oscillation term $\widetilde{\eta }c^{\ast }$
effectively vanishes by averaging. In those cases the LLG dynamics can be
used as an approximation.

The most important influence of this parasitic fast oscillation appears in
the case when we study the effect of a random force $f(t)$ in (\ref{LinLL});
this term gives effectively higher magnetization noise in the system.\cite
{BertJinSaf},\cite{SafBertNoise}. The thermal bath produces both
magnetization relaxation and thermal fluctuations whose level is defined by
fluctuation-dissipation theorem. Both damping and thermal noise appear as a
result of microscopic spin fluctuations, therefore a connection of
phenomenological dynamic magnetization equations with microscopic physics is
necessary. We can quote van Kampen \cite{Kampen}: ``The moral is that one
cannot give a stochastic description of a physical system simply by taking
one of the familiar equations for the sure case and declaring some of the
coefficients to be random. One has to start from the fundamental
equations...''.

\section{Analysis of FMR experimental data}

FMR linewidth measurements have been widely utilized to determine the
relaxation rate for low-level linear excitations \cite{Pattonbook}. In this
section we will use the relaxation rates for microscopic mechanisms analyzed
in Secs. III and IV to analyze FMR data in soft thin films (NiFe, permalloy).

For small relaxation compared to the resonant frequency the linewidth is
given by (e.g., \cite{BertJinSaf}): 
\begin{equation}
\triangle \omega =2\eta  \label{freqline}
\end{equation}
for all microscopic processes. Usually in FMR experiments, the field swept
linewidth $\Delta H$ is measured. There is a simple relation between $\Delta
H$ and frequency linewidth $\Delta \omega $:

\begin{equation}
\Delta H=\frac{\Delta \omega }{\partial \omega _{0}/\partial H_{0}},
\label{fieldline}
\end{equation}
where from (\ref{FMRfreq})

\begin{equation}
\partial \omega _{0}/\partial H_{0}=\gamma ^{2}(H_{x_{0}}+H_{y_{0}})/2\omega
_{0},  \label{domegadh}
\end{equation}
which is valid for any direction of the applied magnetic field, including
those parallel and perpendicular to the film plane.

Let us summarize the linewidths obtained in Sections III\ and IV. Using Eqs.
(\ref{estimatekampat}), (\ref{final}), (\ref{fieldline}) and (\ref{domegadh}%
), we can obtain the field swept linewidths: 
\[
\text{{\it magnon-electron confluence process}} 
\]

\begin{equation}
\Delta H_{m-e}\simeq \frac{c_{def}}{4\pi ^{3}}\frac{\omega _{0}}{\gamma }%
\frac{m^{3}V_{0}^{2}|f|^{2}\epsilon _{F}}{\hbar ^{4}},  \label{fieldkampat}
\end{equation}
\[
\text{{\it slow relaxing impurity process}} 
\]

\begin{equation}
\Delta H_{sr}\simeq \frac{4c_{imp}S}{\hbar \gamma }\frac{(z_{imp}B)}{k_{B}T}%
^{2}\frac{\exp (\hbar \Omega _{0}/k_{B}T)}{[\exp (\hbar \Omega
_{0}/k_{B}T)+1]^{2}}\frac{\omega _{0}\Gamma _{\parallel }}{\Gamma
_{\parallel }^{2}+\omega _{0}^{2}}.  \label{fieldfinal}
\end{equation}
Note that the linear frequency dependence for $\Delta H_{m-e}$ was predicted
in Ref.\cite{KamberskyPatton}.

The temperature dependence of $\Delta H$ in permalloy thin films with the
static magnetic field in the film plane has been measured by Patton and
Wilts \cite{PattonWilts}. As seen in Fig.2, $\Delta H$ exhibits a strong
frequency dependence with a maximum in the vicinity of $T_{\max }\approx 80$
K. The temperature maximum shifts to slightly higher temperatures with
increasing FMR frequency. Such a non-monotonic temperature dependence of $%
\Delta H$ is typical for slow relaxing impurities (see, \cite{PattonWilts}, 
\cite{gurevich},\cite{tchernev}). However, the slow-relaxing impurity
process alone can not describe the experiment \cite{PattonWilts}. One can
assume that the magnon-electron confluence process plays a role of a
relaxation rate ``baseline''. Thus for analysis we combine the slow-relaxing
impurity and magnon-electron confluence processes together:

\begin{equation}
\Delta H=\Delta H_{sr}(T,\omega _{0},c_{imp})+\Delta H_{m-e}(\omega
_{0},c_{def}).  \label{fits}
\end{equation}
According to Eq.(\ref{fieldfinal}), the temperature $T_{\max }$ of linewidth
maximum is given by

\begin{equation}
\Gamma _{\parallel }(T_{\max })\simeq \omega _{0}.  \label{tempmax}
\end{equation}
The impurity damping $\Gamma _{\parallel }$ is expected to increase
monotonically with temperature \cite{gurevich}. Thus $T_{\max }$ will
increase with increasing $\omega _{0}$ in agreement with experiment. The
solid lines in Fig.2 represent a theoretical fit with Eq.(\ref{fits}). We
see that the temperature (and frequency) dependence (\ref{fits}) describe
well the range from about the peak to about room temperature.

For the slow relaxation mechanism the fit gives

\begin{equation}
\Delta H_{sr}[\text{Oe}]\simeq \frac{1.9\cdot 10^{4}}{T[{\rm K}]}\frac{\exp
\left( \frac{100}{T[{\rm K}]}\right) }{\left[ \exp \left( \frac{100}{T[{\rm K%
}]}\right) +1\right] ^{2}}\frac{\frac{\omega _{0}}{\Gamma _{\parallel }}}{%
1+\left( \frac{\omega _{0}}{\Gamma _{\parallel }}\right) ^{2}},
\label{fittemperature}
\end{equation}
where typical parameters for slow-relaxing impurity were used: $\hbar \Omega
_{0}/k_{B}=100$ K, $z_{imp}B/k_{B}=25.3$ K, $S\simeq 1$ and $c_{imp}=10^{-3}$%
. The spin-lattice relaxation rate variation with temperature was assumed to
be $\Gamma _{\parallel }(T)=c_{\parallel }T^{2}$ with a fit of $c_{\parallel
}/2\pi =0.8\cdot 10^{-3}$ GHz/K$^{2}$. The impurity interactions with
conduction electrons are likely to be responsible for such a temperature
dependence. The spin-lattice relaxation time at room temperature $1/\Gamma
_{\parallel }(300$ K$)\simeq 2.2$ psec agrees with typical experimental data 
\cite{gurevich}.

For the magnon-electron confluence process the fit gives

\begin{equation}
\Delta H_{m-e}[{\rm Oe}]\simeq 2.1\left( \omega _{0}/2\pi \lbrack {\rm GHz}%
]\right) .  \label{m-e process}
\end{equation}
Substituting $m\simeq 10^{-27}$ g, $V_{0}\simeq 10^{-22}$ cm$^{3}$ and $%
\gamma =1.76\ 10^{7}$ s$^{-1}$Oe$^{-1}$ into Eq.(\ref{fieldkampat}) and
comparing with Eq.(\ref{m-e process}), we obtain $c_{def}|\hbar
f|^{2}\epsilon _{F}\simeq 7\cdot 10^{-38}$ erg$^{3}$. Taking defect
concentration $c_{def}=10^{-3}$ and Fermi energy $\epsilon _{F}=3$ eV, one
gets an estimate for the interaction amplitude $\hbar f\simeq 2.4$ eV, which
is typical for ferromagnetic metals \cite{krinchik}.

Patton et al \cite{PattonFraitWilts}\ have studied the frequency dependence
of $\Delta H$ in thin permalloy films (17-48 nm) at room temperature (see,
Fig.3). They have found that the field linewidth has a linear frequency
dependence $\Delta H\propto \omega _{0}$ for the case when the external
magnetic field is parallel to the film plane (open circles in Fig.2). For
the perpendicular case (black circles) the linear dependence is valid just
for high FMR frequencies $\omega _{0}/2\pi >8$ GHz and it is saturated at
smaller frequencies. Recent data \cite{michmic} show similar results except
that for the field perpendicular to the film plane the saturation region
occurs only at extremely small frequencies ($\omega _{0}/2\pi <2$ GHz).
Essentially for a wide range of frequencies the field swept linewidths are
identical for the two field orientations and strictly linear with resonant
frequency. It is clear that this linearity can be explained by the
magnon-electron confluence process (\ref{m-e process}). At room temperature
the impurity relaxation rate is $\Gamma _{\parallel }/2\pi \simeq 72$ GHz so
that this mechanism yields a linewidth (\ref{fittemperature}) that is also
linear with frequency at least to about 50 GHz. Taking (\ref{m-e process}), (%
\ref{fittemperature}) and (\ref{fits}) at $T=300$ K we obtain:

\begin{equation}
\Delta H[{\rm Oe}]\simeq 2.3\left( \omega _{0}/2\pi \lbrack {\rm GHz}%
]\right) .  \label{frequencyfit}
\end{equation}
This linear dependence is shown in Fig.3 by a solid line and demonstrates
good agreement with experiment.

It is interesting that Eq.(\ref{fits}) can be also used for interpretation
of experimental field linewidth in ultra-thin permalloy films \cite{azevedo}%
. The thickness dependence at room temperature for two different FMR
frequencies is shown in Fig.4. Assuming that the surface magnetic atoms
behaves as impurities, the concentration $c_{imp}$ increases with decreasing
film thickness as $c_{imp}\propto d_{0}/d$, where $d_{0}\sim 3$ \AA\ is the
linear size of the single-atomic magnetic layer. The theoretical thickness
dependence for typical microscopic parameters (as above) can be written as

\begin{equation}
\Delta H[{\rm Oe}]=\left( 1.3+53/d[\text{\AA }]\right) \left( \omega
_{0}/2\pi \lbrack {\rm GHz}]\right) .  \label{azevedo}
\end{equation}
The solid lines in Fig.4 by plot Eq.(\ref{azevedo}) and demonstrate good
agreement with experimental data. This is an alternative interpretation of 
\cite{azevedo}, where two-magnon scattering on surface defects (with
thickness dependence $1/d^{2}$) \cite{AriasMills} plus a baseline thickness
independent process were suggested. In order to evaluate these two
interpretations, FMR linewidth should be measured at lower temperatures,
where the role of slow-relaxing impurities increases.

\subsection{Comment: data fitting by the LLG equation}

The Landau-Lifshitz-Gilbert equations were originally introduced (a) for
small magnetization motions (small deviations from the equilibrium) and (b)
for the case of high magnetic symmetry (axial symmetry) with an isotropic
damping fitting parameter $\alpha $ (or ``damping constant'') \cite{landau}, 
\cite{gilbert}. Nevertheless both ``a'' and ``b'' conditions are usually
violated; these equations have been used for large magnetization motions
such as magnetization reversal and for a description of magnetization
dynamics in systems with lower magnetic symmetry (e.g., in thin films).

In the LLG phenomenology the linewidth is given by (e.g., \cite{BertJinSaf}):

\begin{equation}
\triangle \omega =\alpha \gamma \left( H_{x_{0}}+H_{y_{0}}\right) .
\label{LLGfreqlinewidth}
\end{equation}
From Eqs. (\ref{domegadh}) and (\ref{fieldline}) we find that the
corresponding field swept linewidth is: 
\begin{equation}
\triangle H=\alpha \omega _{0}/\gamma .  \label{LLGfieldline}
\end{equation}
This formula has been used to obtain the phenomenological damping parameter $%
\alpha $ from experimental data $\triangle H$ and frequency $\omega _{0}$.

Assuming a constant $\alpha $ leads, according to (\ref{LLGfieldline}), to a
linear frequency dependence for $\triangle H$, in agreement with the room
temperature data shown in Fig.3. This agreement has been used to claim the
validity of the LLG equations even for an anisotropic system, e.g. a thin
film \cite{KamberskyPatton}. This dependence is obviously coincidental. The
LLG approach is purely phenomenological and can be related to physical
processes only in the case of axial symmetry. As we have shown in this
paper, detailed microscopic analysis yields dynamic equations which depend
on the system symmetry. Microscopic analysis with appropriate relaxation
mechanisms explain a wide variety of experimental data and, as well, can be
used to predict dynamic magnetization phenomena.

\section{Discussion}

In this paper we have examined the effect of general forms of microscopic
relaxation mechanisms on magnetization dynamics. Only the case of linear
relaxation has been considered when the deviations of magnetization from
equilibrium are relatively small. We have shown for two broad classes of
relaxation mechanisms that dynamic relaxation may be expressed in terms of a
damped harmonic oscillator equation driven by thermal fluctuations. The
coordinates of the harmonic oscillator reflect a transformation of the
transverse magnetization coordinates to circular motion. The transformation
directly reflects the magnetic anisotropy. The relaxation time or damping
parameter contains details of the microscopic relaxation mechanisms and
also, in general, information, via the transformation, about the magnetic
anisotropy. It is important to note that by anisotropy we do not mean the
for of the interaction with the thermal bath. Here the anisotropy that
yields a tensor form is macroscopic and arises, for example, from the sample
shape (e.g., a thin film) or the crystalline symmetry.

Specific mechanisms of slow relaxing impurities and magnon-electron coupling
were analyzed. It was shown that a sum of these two effects explains very
well FMR field swept linewidth measurements versus temperature, frequency
and film thickness. The fitting parameters agree well with previous
estimates.

Transformation of the harmonic oscillator back to magnetization components
results in a macroscopic form that is different from that of LLG. The
damping term, although proportional to the relaxation time or damping
parameter, depends on the magnetization and effective field through a
tensor, whose components reflect the magnetic anisotropy. LLG results only
for the case of uniaxial system anisotropy about equilibrium. The LLG
equations for a fixed damping parameter $\alpha $ do show a linear frequency
dependence of the linewidth, in agreement with data. However. this is just
an artifact: the LLG damping can not be derived in terms of a microscopic
relaxation rate.

In \cite{BertSafJin} and earlier in \cite{safbertnonuni} it was shown that a
magnetic system capable of non-uniform excitations about (a possibly)
non-uniform equilibrium also resulted in damped harmonic oscillators. In
this general case the eigenmodes of the general gyromagnetic motion tensor
(for a discretized system) yield independent harmonic oscillators with
corresponding resonance frequencies, relaxation rates and stochastic thermal
energy. Examples were given for a thin film sensor in a GMR transducer and
in \cite{JinBertH} experimental fitting showed a better agreement than with
simply adding energy to each discretized computation cell and applying LLG
with random fields to each cell. Note that the dynamic normal mode approach
with phenomenologically introduced damping\ (as a small imaginary addition
to the frequency) has been successfully used in the theory of parametrically
excited spin waves (see, e.g., \cite{Suhl}, \cite{Schloemann},\cite
{Pattonbook}).

We have examined microscopic magnetization dynamics for two broad classes of
interactions with a thermal bath. Further work will study the role of
anisotropy in the case of magnetic multilayers \cite{heinrich} and other
indirect processes \cite{SuhlJAP}.

\section{Acknowledgments}

The authors would like to thank H. Suhl, C. E. Patton, V. Kambersky, and R.
D. McMichael for helpful discussions. This work was partly supported by the
Center for Magnetic Recording Research at the University of California - San
Diego and a research gift from the Seagate Corporation.

\appendix 

\section{Elliptic magnetization motion}

Consider the following energy density: 
\begin{equation}
{\cal E}_{p}/V=K_{\parallel }\sin ^{2}\theta +K_{\perp }(M_{x}/M_{s})^{2}-%
{\bf M\cdot H}_{0}.  \label{ex_energy}
\end{equation}
Here $K_{\parallel }\geq 0$ and $K_{\perp }\geq 0$\ are the ``easy'' ($z$)
and ``hard'' ($x$) uniaxial anisotropy constants, $\theta $ is the angle
between the magnetization ${\bf M=}(M_{x},M_{y},M_{z})\;$and the easy axis
of anisotropy. ${\bf H}_{0}=H_{0}(0,~\sin \theta _{H},~\cos \theta _{H})$ is
the external magnetic field. The equilibrium magnetization state $\theta
=\theta _{0}$\ is defined by the condition $\,\partial {\cal E}_{p}/\partial
\theta =0$, or,

\begin{equation}
H_{{\rm K}}^{(\parallel )}\sin 2\theta _{0}=2H_{0}\sin (\theta _{H}-\theta
_{0}),  \label{equi}
\end{equation}
where $\,H_{{\rm K}}^{(\parallel )}\equiv 2K_{\parallel }/M_{s}$. In order
to describe small magnetization oscillations it is necessary to transform
the coordinate system and magnetization as follows

\begin{eqnarray}
y &=&y_{0}\cos \theta _{0}+z_{0}\sin \theta _{0},  \label{yztransform} \\
z &=&-y_{0}\sin \theta _{0}+z_{0}\cos \theta _{0},\quad x=x_{0},  \nonumber
\end{eqnarray}
and

\begin{eqnarray}
M_{y} &=&M_{y_{0}}\cos \theta _{0}+M_{z_{0}}\sin \theta _{0},
\label{yztransform1} \\
M_{z} &=&-M_{y_{0}}\sin \theta _{0}+M_{z_{0}}\cos \theta _{0},\quad
M_{x}=M_{x_{0}}.  \nonumber
\end{eqnarray}
The equilibrium state in these coordinates is ${\bf M=}(0,\ 0,\ M_{s})$.
Taking into account transformation (\ref{yztransform1}) and relations $%
M_{z_{1}}^{2}=M_{s}^{2}-M_{x_{0}}^{2}-M_{y_{0}}^{2}$ and $M_{z_{0}}\simeq
M_{s}-(M_{x_{0}}^{2}+M_{y_{0}}^{2})/2M_{{\rm s}}$, the energy (\ref
{ex_energy}) can be written in the form (\ref{energy}), where 
\begin{eqnarray}
H_{x_{0}} &=&H_{{\rm K}}^{(\parallel )}\cos ^{2}\theta _{0}+H_{{\rm K}%
}^{(\perp )}+H_{0}\cos (\theta _{H}-\theta _{0}),  \nonumber \\
H_{y_{0}} &=&H_{{\rm K}}^{(\parallel )}\cos 2\theta _{0}+H_{0}\cos (\theta
_{H}-\theta _{0}).  \label{H1}
\end{eqnarray}
and $H_{{\rm K}}^{(\perp )}\equiv 2K_{\perp }/M_{s}$. From (\ref{H1}) and (%
\ref{equi}) it follows that $H_{x_{0}}=H_{y_{0}}$ only in two special cases:
1) the case of spherical symmetry and 2) the case of uniaxial symmetry when
the external magnetic field and equilibrium magnetization are oriented along
the easy axis.

\section{Magnon-electron scattering without momentum conservation}

The interaction of the uniform magnetic precession ($a$, $a^{\ast }$) with
electrons is assumed to be of the form:

\begin{equation}
{\cal H}_{I}=\left( D^{\dagger }a+Da^{\ast }\right) /\sqrt{N},  \label{KP6}
\end{equation}
where

\begin{equation}
D=\frac{1}{N}\sum_{{\bf k},{\bf k}^{\prime }}\sum_{j}f_{{\bf kk}^{\prime }}(%
{\bf r}_{j})d_{{\bf k}^{\prime }}^{\dagger }d_{{\bf k}}.  \label{KP12}
\end{equation}
Here $f_{{\bf kk}^{\prime }}({\bf r}_{j})=|f_{{\bf kk}^{\prime }}|\exp
[i\phi ({\bf r}_{j})]$ describes the scattering process in the vicinity of
crystal defect at the point ${\bf r}_{j}$, $\phi ({\bf r}_{j})$ is the
phase, $d_{{\bf k}}^{\dagger }$ and $d_{{\bf k}}$ are the fermion operators
describing creation and annihilation of electron with wave number ${\bf k}$
and frequency $\omega _{e{\bf k}}$.

Using transformation (\ref{canonic}) to a normal mode, we can rewrite ${\cal %
H}_{I}$ as:

\begin{equation}
{\cal H}_{I}/\hbar =\frac{1}{N^{3/2}}\sum_{{\bf k},{\bf k}^{\prime }}\sum_{j}%
\left[ \Psi _{{\bf kk}^{\prime }}({\bf r}_{j})d_{{\bf k}^{\prime }}^{\dagger
}d_{k}c+\Psi _{{\bf kk}^{\prime }}^{\ast }({\bf r}_{j})d_{{\bf k}}^{\dagger
}d_{{\bf k}^{\prime }}c^{\ast }\right] ,  \label{SB4}
\end{equation}
where

\begin{equation}
\Psi _{{\bf kk}^{\prime }}({\bf r}_{j})=uf_{{\bf kk}^{\prime }}({\bf r}%
_{j})+vf_{{\bf k}^{\prime }{\bf k}}^{\ast }({\bf r}_{j})  \label{Psi}
\end{equation}
is the amplitude of magnon-electron scattering.

The relaxation rate is defined by the golden Fermi rule as 
\begin{eqnarray}
\eta _{m-e} &=&\frac{\pi }{N^{3}}\sum_{{\bf k},{\bf k}^{\prime
}}\sum_{j}|\Psi _{{\bf k}^{\prime }{\bf k}}|^{2}(\overline{n}_{{\bf k}}-%
\overline{n}_{{\bf k}^{\prime }})\delta (\omega _{e{\bf k}^{\prime }}-\omega
_{e{\bf k}}-\omega _{0})  \nonumber \\
&=&\frac{\pi }{N^{3}}\sum_{{\bf k},{\bf k}^{\prime }}\sum_{j}\{u^{2}|f_{{\bf %
k}^{\prime }{\bf k}}|^{2}+v^{2}|f_{{\bf kk}^{\prime }}|^{2}  \nonumber \\
&&+uv|f_{{\bf k}^{\prime }{\bf k}}f_{{\bf kk}^{\prime }}|\cos [2\phi ({\bf r}%
_{j})]\}  \nonumber \\
&&\times (\overline{n}_{{\bf k}}-\overline{n}_{{\bf k}^{\prime }})\delta
(\omega _{e{\bf k}^{\prime }}-\omega _{e{\bf k}}-\omega _{0}).
\label{oneovertau}
\end{eqnarray}
Here $\overline{n}_{{\bf k}}\equiv \langle d_{{\bf k}}^{\dagger }d_{{\bf k}%
}\rangle $ is the Fermi occupation number If we assume that the scattering
phases $\phi ({\bf r}_{j})$ are random and $|f_{{\bf k}^{\prime }{\bf k}%
}|=|f_{{\bf kk}^{\prime }}|$, then the $uv$ term vanishes and\ Eq.(\ref
{oneovertau}) becomes

\begin{eqnarray}
\eta _{m-e} &=&\frac{c_{def}\pi (u^{2}+v^{2})}{N^{2}}\sum_{k,k^{\prime }}|f_{%
{\bf kk}^{\prime }}|^{2}  \nonumber \\
&&\times (\overline{n}_{{\bf k}}-\overline{n}_{{\bf k}^{\prime }})\delta
(\omega _{e{\bf k}^{\prime }}-\omega _{e{\bf k}}-\omega _{0}).  \label{KP14}
\end{eqnarray}
Here $c_{def}=N_{def}/N$, $N_{def}$ is the total number of defects. After
the following simplifications: $\overline{n}_{{\bf k}}-\overline{n}_{{\bf k}%
^{\prime }}\simeq -\omega _{0}\partial \overline{n}_{{\bf k}}/\partial
\omega _{e{\bf k}}$ and $\omega _{0}\ll \omega _{0}$, we obtain Eq.(\ref
{kampat}).

\section{Interactions with impurities}

Let us consider two-level impurities as effective spins $s_{j}=1/2$. The
anisotropic exchange Hamiltonian between the impurities and the neighboring
host spins can be written in a general form:

\begin{equation}
{\cal H}_{{\rm ex}}=\sum\limits_{j,\nu }\sum_{a_{j},a}B_{a_{j},a}({\bf R}%
_{j},{\bf r}_{\nu })s_{a_{j}}{({\bf R}_{j})}S_{a}({\bf R}_{j}+{\bf r}_{\nu
}).  \label{Ham-imp}
\end{equation}
Here $B_{a_{j},a}({\bf R}_{j},{\bf r}_{\nu })$ are the anisotropic exchange
integrals,$\ a_{j}=x_{j},y_{j},z_{j}$ are the local principal axes for the $%
j $-th impurity ``spin'' ${\bf s}_{j}({\bf R}_{j})=\{s_{x_{j}}({\bf R}_{j}),{%
s_{y_{j}}({\bf R}_{j}),s_{z_{j}}({\bf R}_{j})}\}$ located at ${\bf R}_{j}$,
and ${\bf S}({\bf R}_{j}+{\bf r}_{\nu })=\{S_{x_{0}}({\bf R}_{j}+{\bf r}%
_{\nu }),S_{y_{0}}({\bf R}_{j}+{\bf r}_{\nu }),S_{z_{0}}({\bf R}_{j}+{\bf r}%
_{\nu })\}$ are the\ components of the host spin located in the vicinity of $%
j$-th impurity.

The Hamiltonian describing the impurity level modulation follows from Eq.(%
\ref{Ham-imp}):

\begin{eqnarray}
{\cal H}_{ex,slow} &=&\frac{1}{\hbar }\sum\limits_{j,\nu }{\big[}%
B_{z_{j},x_{0}}({\bf R}_{j},{\bf r}_{\nu })S_{x_{0}}({\bf R}_{j}+{\bf r}%
_{\nu })  \label{imp-ex} \\
&&+B_{z_{j},y_{0}}({\bf R}_{j},{\bf r}_{\nu })S_{y_{0}}({\bf R}_{j}+{\bf r}%
_{\nu })\big]{s_{z_{j}}({\bf R}_{j}).}  \nonumber
\end{eqnarray}

\newpage

{\large Figure captions}

Fig.1 Relaxation processes: a) confluence of magnon and conduction electron
into conduction electron in the presence of defect, b) `slow-relaxation'
mechanism.via modulation of impurity levels by magnetization oscillation.

Fig.2 The linewidth temperature dependence for three FMR frequencies. Points
are experimental data of Ref. \cite{PattonWilts}. Solid lines demonstrate
theoretical fit with (\ref{fittemperature}) and (\ref{m-e process}).

Fig.3 FMR frequency dependence at room temperature. Open circles correspond
to parallel and black point to perpendicular field orientations \cite
{PattonFraitWilts}. Solid line demonstrate theoretical fit (\ref
{frequencyfit}).

Fig.4 The linewidth thickness dependence for two FMR frequencies at room
temperature \cite{azevedo}. Solid lines demonstrate theoretical fit (\ref
{azevedo}).


\begin{references}
\bibitem{landau}  L. Landau and E. Lifshitz, Phys. Z. Sowjet. {\bf 8}, 153
(1935); in Landau L. D. {\it Collected Papers}. Edited by D. ter Haar
(Gordon and Breach, New York, 1967) p. 101.

\bibitem{gilbert}  T. L. Gilbert, Phys. Rev. {\bf 100}, 1243 (1955); T. L.
Gilbert ``Formulation, Foundations and Applications of the Phenomenological
Theory of Ferromagnetism'', Ph.D.thesis, Illinois Institute of Technology,
Chicago, Illinois, June 1956, unpublished.

\bibitem{callen}  H. B. Callen, J. Phys. Chem. Solids {\bf 4}, 256 (1958).

\bibitem{Patton}  C. E. Patton, J. Appl. Phys. {\bf 39}, 3060 (1968).

\bibitem{baryakhtar}  V. G. Bar'yakhtar, Sov. Phys. Solid State {\bf 29},
754 (1987); Physica B {\bf 159}, 20 (1989).

\bibitem{Silvabook}  T. J. Silva, ``Measurement of Dynamic Properties in
Thin-Films'', in {\it The Physics of Ultra-High-Density Magnetic Recording},
edited by M. Plumer, J. van Ek, and D. Weller (Springer, Berlin, 2001),
p.110.

\bibitem{BertSafJin}  H. N. Bertram, V. L. Safonov, and Z. Jin, IEEE Trans.
Magn. {\bf 38}, \#5, Septermber (2002) in press.

\bibitem{SafTensor}  V. L. Safonov, J. Appl. Phys. {\bf 91}, 8653 (2002).

\bibitem{BertJinSaf}  H. N. Bertram, Z. Jin, and V. L. Safonov, IEEE Trans.
Magn. {\bf 38}, 38 (2002).

\bibitem{SafBertNoise}  V. L. Safonov and H. N. Bertram, Phys. Rev. B {\bf 65%
}, 172417 (2002).

\bibitem{PattonWilts}  C. E. Patton and C. H. Wilts, J. Appl. Phys. {\bf 38}%
, 3537 (1967).

\bibitem{PattonFraitWilts}  C. E. Patton, Z. Frait, and C. H. Wilts, J.
Appl. Phys. {\bf 46}, 5002 (1975).

\bibitem{sparks}  M. Sparks, {\it Ferromagnetic-Relaxation Theory }%
(McGraw-Hill, New York, 1964).

\bibitem{ABP}  A. I. Akhiezer, V. G. Baryakhtar and S. V. Peletminskii, {\it %
Spin Waves }(North Holland, Amsterdam, 1968).

\bibitem{hopri}  T. Holstein and H. Primakoff, Phys. Rev. {\bf 58}, 1098
(1940).

\bibitem{Lax0}  M. Lax, Phys. Rev. {\bf 145}, 110 (1966).

\bibitem{lax}  M. Lax, in {\it Statistical Physics, Phase Transitions, and
Superconductivity}, edited by M. Chr\'{e}tien, E. P. Gross, and S. Des\'{e}r
(Gordon and Breach, New York, 1968), Vol.2, p.269.

\bibitem{Louisell}  W. H. Louisell, {\it Quantum Statistical Properties of
Radiation} (Willey, New York, 1973).

\bibitem{scully}  M. O. Scully and M. S. Zubairy, {\it Quantum optics}
(Cambridge University Press, 1997).

\bibitem{twomagnon}  M. J. Hurben and C. E. Patton, J. Appl. Phys. {\bf 83},
4344 (1998).

\bibitem{AriasMills}  R. Arias and D. L. Mills, Phys. Rev. B {\bf 60}, 7395
(1999).

\bibitem{mcmichael}  R. D. McMichael and A. Kunz, J. Appl. Phys. {\bf 91},
8650 (2002).

\bibitem{reizer}  V. S. Lutovinov and M. Yu. Reizer, Sov. Phys. JETP {\bf 50}%
, 355 (1979).

\bibitem{solontsov}  A. Z. Solontsov and A. N. Vasil'ev, Phys. Lett. A 177,
362 (1993); J. Magn. Magn. Mater. 140-144, 213 (1995).

\bibitem{KamberskyPatton}  V. Kambersky and C. E. Patton, Phys. Rev. B {\bf %
11}, 2668 (1975).

\bibitem{mikhailov}  A. S. Mikhailov and R. M. Farzetdinova, Sov. Phys. JETP 
{\bf 53}, 782 (1981); {\it ibid.} {\bf 57}, 109 (1983).

\bibitem{saffar}  V. L. Safonov and R. M. Farzetdinova, Phys. stat. solidi
(b) {\bf 163}, 259 (1991).

\bibitem{safbertimpurity}  V. L. Safonov and H. N. Bertram, Phys. Rev. B 
{\bf 61}, R14893 (2000).

\bibitem{gurevich}  A. G. Gurevich and G. A. Melkov, {\it Magnetization
Oscillations and Waves} (CRC Press, Boca Raton, Fla., 1996).

\bibitem{slownoise}  Random noise in this equation appears from the thermal
fluctuation of $n_{T}$.

\bibitem{Kampen}  N. G. van Kampen, Phys. Reports {\bf 24}, 171 (1976).

\bibitem{Pattonbook}  C. E. Patton, ``Microwave Resonance and Relaxation'',
in {\it Magnetic Oxides}, edited by D. J. Craik (Wiley, London, 1975), p.575.

\bibitem{tchernev}  D. I. Tchernev, J. Appl. Phys. {\bf 38}, 1046 (1967).

\bibitem{krinchik}  G. S. Krinchik, {\it Physics of Magnetic Phenomena }%
(Moscow State University, Moscow, 1085) in Russian.

\bibitem{michmic}  R. D. McMichael,\ private communication.

\bibitem{azevedo}  A. Azevedo, A. B. Oliveira, F. M. de Aguiar, and S. M.
Rezende, Phys. Rev. B {\bf 62}, 5331 (2000).

\bibitem{safbertnonuni}  V. L. Safonov and H. N. Bertram, J. Appl. Phys. 
{\bf 91}, 7279 (2002).

\bibitem{JinBertH}  Z. Jin, H. N. Bertram, and H. Dakroub, IEEE Trans. Magn. 
{\bf 38}, \#5, Septermber (2002) in press.

\bibitem{Suhl}  H. Suhl, J. Phys. Chem. Solids {\bf 1}, 215 (1957).

\bibitem{Schloemann}  E. Schlomann, R. I. Joseph, and I. Bady, J. Appl. Phys.%
{\bf \ 34}, 672 (1963).

\bibitem{heinrich}  E. Sim\'{a}nek and B. Heinrich, arXiv: cond-mat/
0207471, Jul 2002.

\bibitem{SuhlJAP}  H. Suhl, J. Appl. Phys. {\bf 89}, 7448 (2001).
\end{references}
\end{document}